# Regional Climate Change Datasets for South Asia

J. Sanjay, M. V. S. Ramarao, R. Mahesh, Sandip Ingle, BhupendraBahadur Singh, R. Krishnan

Centre for Climate Change Research, Indian Institute of Tropical Meteorology, Pune 411008

Email: sanjay@tropmet.res.in

## 1. Introduction

The Centre for Climate Change Research (CCCR; http://cccr.tropmet.res.in) at the Indian Institute of Tropical Meteorology (IITM; http://www.tropmet.res.in), Pune, launched in 2009 with the support of the Ministry of Earth Sciences (MoES), Government of India, focuses on the development of new climate modelling capabilities in India and South Asia to address issues concerning the science of climate change. CCCR-IITM has the mandate of developing an Earth System Model and to make the regional climate projections. An important achievement was made by developing an Earth System Model at IITM, which is an important step towards understanding global and regional climate response to long-term climate variability and climate change. CCCR-IITM has also generated an ensemble of high resolution dynamically downscaled future projections of regional climate over South Asia and Indian monsoon, which are found useful for impact assessment studies and for quantifying uncertainties in the regional projections. A brief overview of these core climate change modeling activities of CCCR-IITM was presented in an Interim Report on Climate Change over India (available at http://cccr.tropmet.res.in/home/reports.jsp)

## 2. Regional Climate Projections for South Asia

The ensemble of high resolution downscaled projections of regional climate and monsoon over South Asia until 2100 are generated by CCCR-IITM using a regional climate model (ICTP-RegCM4; Giorgi et al. 2012) at 50 km horizontal resolution. These high-resolution downscaled projections of regional climate over South Asia are developed as part of the WMO's World Climate Research Programme (WCRP) regional activity Coordinated Regional Climate Downscaling Experiment (CORDEX; http://cordex.org/). The CORDEX aims to foster international partnership in order to produce an ensemble of high-resolution past and future climate projections at regional scale. CCCR-IITM is the nodal agency for coordinating the CORDEX modeling activity in South Asia (https://www.wcrp-climate.org/wcrp-regional-activities/ra-asia). The CCCR-IITM and several international partner institutions have contributed towards generation and evaluation of regional climate simulations for CORDEX South Asia (http://cccr.tropmet.res.in/home/cordexsa_datasets.jsp). This CORDEX dataset is comprised of downscaled climate scenarios for the South Asia region that are derived from the Atmosphere-Ocean coupled General Circulation Model (AOGCM) runs conducted under the Coupled Model Intercomparison Project Phase 5 (CMIP5), and using three of the four greenhouse gas emissions scenarios known as Representative Concentration Pathways (RCPs). The CMIP5 AOGCM runs were developed in support of the Fifth Assessment Report of the Intergovernmental Panel on Climate Change (IPCC AR5). The coarser spatial resolution ranging from 1.0° to 3.8°, and systematic error (called bias) of these AOGCMs limits the examination of possible impacts of climate change and adaptation strategies on a smaller scale. The dynamical downscaling method using high resolution limited area regional climate models (RCMs) utilizes the outputs provided by AOGCMs as lateral boundary condition to provide physically consistent spatiotemporal variations of climatic parameters at spatial scales much smaller than the AOGCMs' grid. The CORDEX South Asia dataset includes dynamically downscaled projections from the 10 models and scenarios for which daily scenarios were produced and distributed under CMIP5. The purpose of these datasets is to provide a set of high resolution (50 km) regional climate change projections that can be used to evaluate climate change impacts on processes that are sensitive to finer-scale climate gradients and the effects of local topography on climate conditions. An initial assessment of the ability of the CORDEX RCMs to simulate the general characteristics of the Indian climate indicated that the geographical distribution of surface air temperature and seasonal precipitation in the present climate for land areas in South Asia is strongly affected by the choice of the RCM and boundary conditions (i.e. driving AOGCMs), and the downscaled seasonal averages are not always improved (Sanjay et al. 2017a). The CORDEX South Asia datasets was recently used for assessing the future climate projections over India in the Interim Report on Climate Change over India (available at http://cccr.tropmet.res.in/home/reports.jsp).

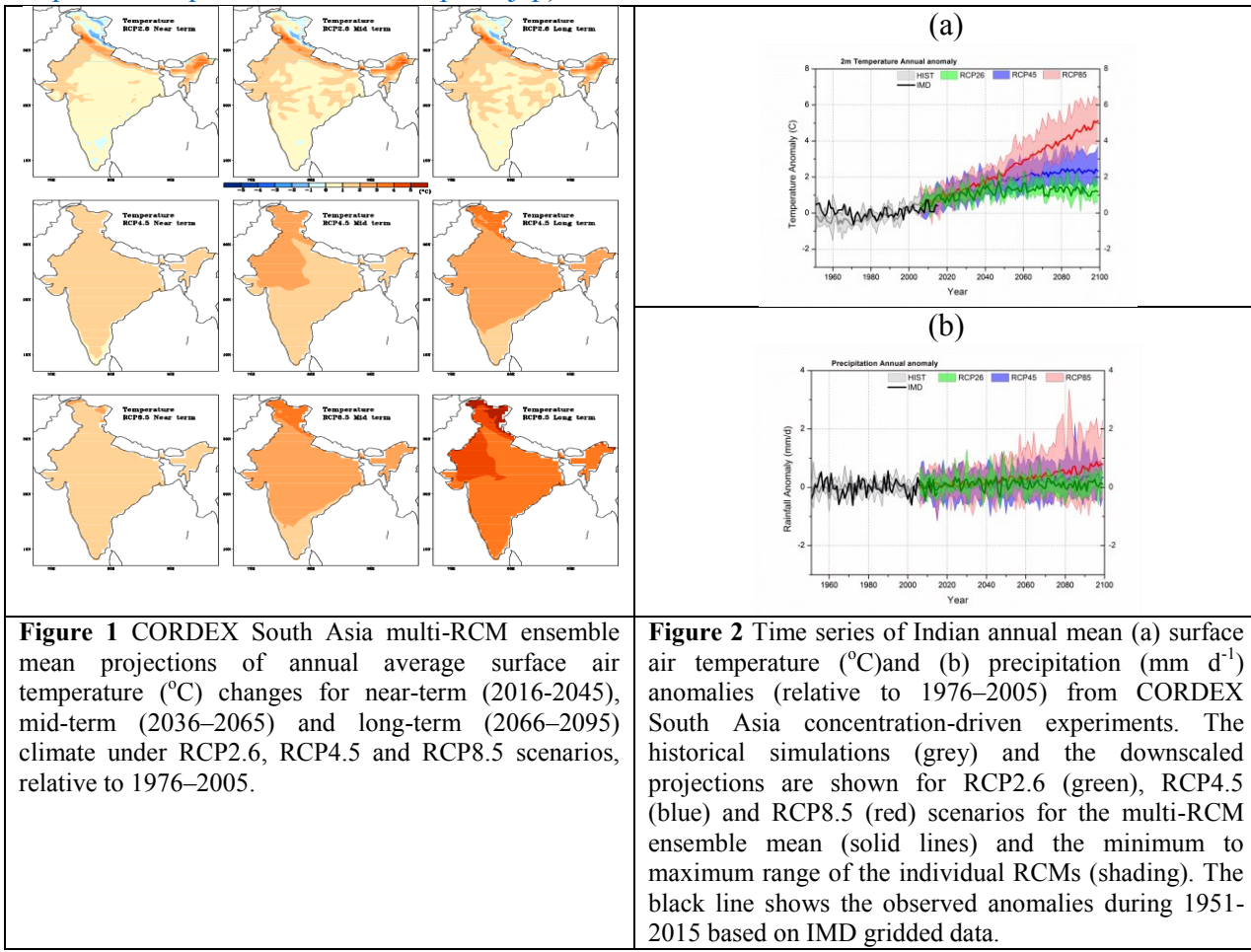

**Figure 1** CORDEX South Asia multi-RCM ensemble mean projections of annual average surface air temperature ($^{o}$C) changes for near-term (2016-2045), mid-term (2036–2065) and long-term (2066–2095) climate under RCP2.6, RCP4.5 and RCP8.5 scenarios, relative to 1976–2005.

**Figure 2** Time series of Indian annual mean (a) surface air temperature ($^{o}$C) and (b) precipitation (mm d$^{-1}$) anomalies (relative to 1976–2005) from CORDEX South Asia concentration-driven experiments. The historical simulations (grey) and the downscaled projections are shown for RCP2.6 (green), RCP4.5 (blue) and RCP8.5 (red) scenarios for the multi-RCM ensemble mean (solid lines) and the minimum to maximum range of the individual RCMs (shading). The black line shows the observed anomalies during 1951-2015 based on IMD gridded data.

The projections of near-term (2016-2045) change in the CORDEX South Asia multi-RCM ensemble mean annual mean surface air temperature relative to the reference period 1976-2005, show modest sensitivity to alternate RCP scenarios over Indian land area (see left panels of Figure 1). The projected annual warming exceeding 3°C is seen over entire India for the high-emission RCP8.5 scenario by the end of 21st century, with relatively higher change exceeding 4°C projected in the semi-arid north-west and north India (see bottom right panel of Figure 1). The CORDEX South Asia historical RCM simulations capture the observed all India averaged annual surface air temperature interannual variations and the warming trend reasonably well (Figure 2a). A consistent and robust feature across the downscaled CORDEX South Asia RCMs is a continuation of warming over India in the 21st century for all the RCP scenarios.The spread in the minimum to maximum range in the projected warming among the CORDEX South Asia RCMs for each RCP scenario (shown as shading in Figure 2a) provide a simple, but crude, measure of uncertainty.

**Table 1** CORDEX South Asia multi-RCM reliability ensemble average (REA) estimate of projected changes in annual mean surface air temperature over India and the associated uncertainty range. The values in parenthesis show the uncertainty in percent for the REA estimate.

| Scenario | Annual Mean Temperature ($^oC$) | | |
|---|---|---|---|
| | 2030s | 2050s | 2080s |
| **RCP2.6** | 1.08 ± 0.12(11.1%) | 1.35 ± 0.18(13.3%) | 1.35 ± 0.23(17.0%) |
| **RCP4.5** | 1.28 ± 0.20(15.6%) | 1.92 ± 0.28(14.6%) | 2.41 ± 0.40(16.6%) |
| **RCP8.5** | 1.44 ± 0.17(11.8%) | 2.41 ± 0.28(11.6%) | 4.19 ± 0.46(11.0%) |

A quantitative estimate of the associated uncertainty range based on a reliability ensemble average (REA) method (Sengupta and Rajeevan 2013) incorporating each RCM performance and convergence indicated that the all India mean surface air temperature is projected to increase in the far future (2066–2095) by 4.19 ± 0.46°C under RCP8.5 scenario, associated with 11% uncertainty range (Table 1). Although the all India annual precipitation is found to increase as temperature increases (Figure 2b), the REA assessment indicates that precipitation changes throughout the 21st century remain highly uncertain (not shown).The new information available from CORDEX South Asia wasalso found useful for contributing to the Hindu Kush Himalayan Monitoring and Assessment Programme (HIMAP; Sanjay et al. 2017b; Krishnan et al. forthcoming).

## 3. Dissemination of Climate Change Datasets

The CORDEX South Asia RCM outputs shared by the modeling partners are archived and published on the CCCR-IITM climate data portal designed to facilitate the dissemination of climate information using a publicly accessible web-based interface. CCCR-IITM has developed an Earth System Grid Federation (ESGF; https://esgf.llnl.gov/) data node for archival, management, retrieval and dissemination of CORDEX South Asia and CMIP6 datasets. ESGF is an international collaboration for the software that powers using a system of geographically distributed peer nodes, most global climate change research, notably assessments by the IPCC. The quality checked CORDEX South Asia RCM outputs at daily, monthly and seasonal time intervals of about 2 terabyte are published on the CCCR-IITM ESGF data node (http://cccr.tropmet.res.in/home/esgf_data.jsp) for dissemination to users and stakeholders.The CORDEX South Asia dataset available on the ESGF can be subset and extracted for downloading using a web based tool developed by CCCR-IITM (http://cccr.tropmet.res.in/home/data_cccrdx.jsp ).This CORDEX dataset for South Asia region is found useful by the science community for evaluating and quantifying uncertainties in future projections at regional-scales,in conducting studies of climate change impacts at regional scales, and to enhance public understanding of possible future climate patterns at the spatial scale of homogenous regions (http://cccr.tropmet.res.in/home/cordexsa_pub.jsp).

**Acknowledgments**

The IITM-RegCM4 simulations were performed using the IITM Aaditya high power computing resources. The Director, IITM is gratefully acknowledged for extending full support to carry out this research work. IITM receives full support from the Ministry of Earth Sciences, Government of India. The World Climate Research Programme's Working Group on Regional Climate, and the Working Group on Coupled Modelling, former coordinating body of CORDEX and responsible panel for CMIP5 are sincerely acknowledged. The climate modelling groups (listed in http://cccr.tropmet.res.in/home/esgf_data.jsp) are sincerely thanked for producing and making available their model output. The Earth System Grid Federation infrastructure (ESGF; http://esgf.llnl.gov/index.html) is also acknowledged.